\documentclass[twoside,twocolumn,9pt]{article}
\usepackage{extsizes}
\usepackage[super,sort&compress,comma]{natbib} 
\usepackage[version=3]{mhchem}
\usepackage[left=1.5cm, right=1.5cm, top=1.785cm, bottom=2.0cm]{geometry}
\usepackage{balance}
\usepackage{mathptmx}
\usepackage{sectsty}
\usepackage{graphicx} 
\usepackage{lastpage}
\usepackage[format=plain,justification=justified,singlelinecheck=false,font={stretch=1.125,small,sf},labelfont=bf,labelsep=space]{caption}
\usepackage{float}
\usepackage{fancyhdr}
\usepackage{fnpos}
\usepackage{caption}
\usepackage{subcaption}
\usepackage[english]{babel}
\addto{\captionsenglish}{%
 
}
\usepackage{array}
\usepackage{droidsans}
\usepackage{charter}
\usepackage[T1]{fontenc}
\usepackage[usenames,dvipsnames]{xcolor}
\usepackage{setspace}
\usepackage[compact]{titlesec}
\usepackage{hyperref}

\usepackage{tabularx} 

\usepackage{epstopdf}

\definecolor{cream}{RGB}{222,217,201}

\begin{document}

\pagestyle{fancy}
\thispagestyle{plain}
\fancypagestyle{plain}{
\renewcommand{\headrulewidth}{0pt}
}

\makeFNbottom
\makeatletter
\renewcommand\LARGE{\@setfontsize\LARGE{15pt}{17}}
\renewcommand\Large{\@setfontsize\Large{12pt}{14}}
\renewcommand\large{\@setfontsize\large{10pt}{12}}
\renewcommand\footnotesize{\@setfontsize\footnotesize{7pt}{10}}
\makeatother

\renewcommand{\thefootnote}{\fnsymbol{footnote}}
\renewcommand\footnoterule{\vspace*{1pt}%
\color{cream}\hrule width 3.5in height 0.4pt \color{black}\vspace*{5pt}} 
\setcounter{secnumdepth}{5}

\makeatletter 
\renewcommand\@biblabel[1]{#1} 
\renewcommand\@makefntext[1]%
{\noindent\makebox[0pt][r]{\@thefnmark\,}#1}
\makeatother 
\renewcommand{\figurename}{\small{Fig.}~}
\sectionfont{\sffamily\Large}
\subsectionfont{\normalsize}
\subsubsectionfont{\bf}
\setstretch{1.125} 
\setlength{\skip\footins}{0.8cm}
\setlength{\footnotesep}{0.25cm}
\setlength{\jot}{10pt}
\titlespacing*{\section}{0pt}{4pt}{4pt}
\titlespacing*{\subsection}{0pt}{15pt}{1pt}

\fancyfoot{}
\fancyfoot[LO,RE]{\vspace{-7.1pt}\includegraphics[height=9pt]{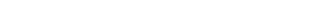}}
\fancyfoot[CO]{\vspace{-7.1pt}\hspace{13.2cm}\includegraphics{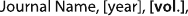}}
\fancyfoot[CE]{\vspace{-7.2pt}\hspace{-14.2cm}\includegraphics{head_foot/RF}}
\fancyfoot[RO]{\footnotesize{\sffamily{1--\pageref{LastPage} ~\textbar \hspace{2pt}\thepage}}}
\fancyfoot[LE]{\footnotesize{\sffamily{\thepage~\textbar\hspace{3.45cm} 1--\pageref{LastPage}}}}
\fancyhead{}
\renewcommand{\headrulewidth}{0pt} 
\renewcommand{\footrulewidth}{0pt}
\setlength{\arrayrulewidth}{1pt}
\setlength{\columnsep}{6.5mm}
\setlength\bibsep{1pt}

\makeatletter 
\newlength{\figrulesep} 
\setlength{\figrulesep}{0.5\textfloatsep} 

\newcommand{\topfigrule}{\vspace*{-1pt}%
\noindent{\color{cream}\rule[-\figrulesep]{\columnwidth}{1.5pt}} }

\newcommand{\botfigrule}{\vspace*{-2pt}%
\noindent{\color{cream}\rule[\figrulesep]{\columnwidth}{1.5pt}} }

\newcommand{\dblfigrule}{\vspace*{-1pt}%
\noindent{\color{cream}\rule[-\figrulesep]{\textwidth}{1.5pt}} }

\makeatother

\twocolumn[
 \begin{@twocolumnfalse}
{\includegraphics[height=30pt]{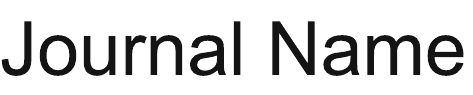}\hfill\raisebox{0pt}[0pt][0pt]{\includegraphics[height=55pt]{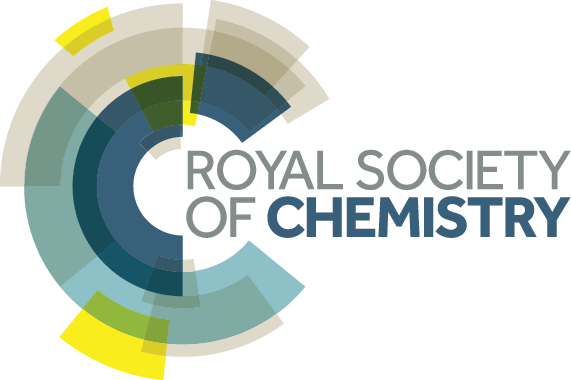}}\\[1ex]
\includegraphics[width=18.5cm]{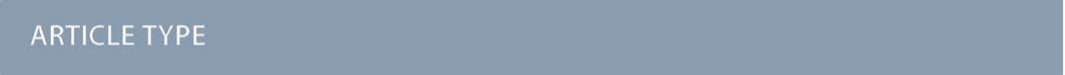}}\par
\vspace{1em}
\sffamily
\begin{tabular}{m{4.5cm} p{13.5cm} }

\includegraphics{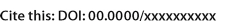} & \noindent\LARGE{\textbf{MXene's Surface Functionalization Patterns and Their Impacts on Magnetism}} \\
\vspace{0.3cm} & \vspace{0.3cm} \\

 & \noindent\large{Barbora Vénosová\textit{$^{a}$} 
 and František Karlický$^{\ast}$\textit{$^{a}$}} \\

\includegraphics{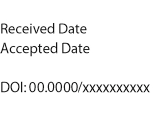} & \noindent\normalsize{}\ 
Two-dimensional transition metal carbides and nitrides (MXenes) are a perspective group of materials with a broad palette of applications. 
Surface terminations are a product of the MXene preparation, and post-processing can also lead to partial coverage. 
Despite applicability and fundamental properties being driven by termination patterns, it is not fully clear, how they behave on MXene surfaces with various degrees of surface coverage. 
Here, as the first step, we used density functional theory to predict possible patterns in prototypic \ce{Ti2C} MXene, demonstrating the different behavior of the two most frequent terminal atoms, oxygen, and fluorine. 
Oxygen (with formal charge -2$e$) prefers a zigzag line both-side adsorption pattern on bare \ce{Ti2C}, attracting the next adsorbent to a minimal distance. 
Oxygen defects in fully O-terminated MXene tend to form similar zigzag line vacancy patterns. 
On the other hand, fluorine (with a formal charge of -1$e$) prefers one-side flake (island) adsorption on bare \ce{Ti2C} and a similar desorption style from fully fluorinated \ce{Ti2C}. 
MXene magnetic behavior is subsequently driven by the patterns, either compensating locally and holding the global magnetic state of the MXene until some limit (oxygen case) or gradually increasing total magnetism by summing of local effects (fluorine case). 
The systematic combinatoric study of \ce{Ti2C}T$_x$ with various coverages ($0 \leq x \leq 2$) of distinct terminal atoms T = O or F brings encouraging possibilities of tunable behavior of MXenes and provides useful guidance for its modeling towards electronic nanodevices.
\end{tabular}
\end{@twocolumnfalse} \vspace{0.6cm}
 ]

\renewcommand*\rmdefault{bch}\normalfont\upshape
\rmfamily
\section*{}
\vspace{-1cm}


\footnotetext{\textit{$^{a}$~Department of Physics, Faculty of Science, University of Ostrava, 30. dubna 22, 7013 Ostrava, Czech Republic. Tel: +420 553 46 2155; E-mail: frantisek.karlicky@osu.cz}}

\footnotetext{\dag~Electronic Supplementary Information (ESI) available: Detailed data for the structures of \ce{Ti2CT2} MXenes: calculated cohesive energy, structure parameters, and Bader charges; spin density distribution of selected MXenes; calculated structures with different surface coverages and patterns of terminal atoms and their relative energies with total magnetic moments; dependence of cohesive energy on the coverage by terminal atoms. See DOI: 00.0000/00000000.}



\section{Introduction}
\label{sec:Intro}
In 2011, a new class of 2D materials named MXenes was discovered\cite{Gogotsi2011} with the general chemical formula M$_{n+1}$X$_n$T$_x$, where M is a transition metal atom (e.g., Ti, V, Sc, Mo, Ta, or Nb), X can be carbon or nitrogen atoms and T stands for the terminal atom/group (which includes groups 16 and 17 of the periodic table or hydroxyl and imido groups). 
In general, MXenes are produced by selectively etching "A" layers from the M$_{n+1}$AX$_n$ phases (where A is mainly the IIIA or IVA group of elements) using an acid solution.\cite{Gogotsi2011, Naguib2012, Naguib2013} 
During the process, the M$_{n+1}$X$_n$ MXenes are completely surface-terminated.\cite{Halim2016,Persson2017}
The content of the attached functional groups depends upon the etching chemicals that have been established experimentally and significantly influence the physical properties of MXenes. \cite{Naguib2013} 
Unlike most 2D layered materials, the functional group can be chemically modified at many locations on the layers of MXenes. 
Thermal and environmental processing can also lead to the removal of a significant number of terminal groups.\cite{Persson2019} 
Taking into account the number of transition metals, carbon, and nitrogen, the four M$_{n+1}$X$_n$T$_x$ (\textit{n} =1 - 4) structures of MXenes, and the known mono-atomic surface terminations, at least a thousand stoichiometric compositions may be possible.\cite{Gogotsi2023} 
The choice of different surface groups for selective termination on MXenes offers remarkable properties and makes them suitable for many potential applications including energy storage, supercapacitors, and photocatalysts. \cite{Naguib2012, Naguib2013} 
Previous studies have shown that surface functionalization is a key factor in modifying the fundamental properties of MXenes and has a decisive influence on the electronic, magnetic, and optical properties of MXenes.\cite{Akgenc2019, Zhang2017}
In this context, several studies have been carried out considering the effect of surface functionalization on electronic properties,\cite{Gao2016, Khazaei2012, Hart2019, Fanta2023} optical properties,\cite{Berdiyorov2016, Xu2020} magnetic,\cite{Champagne2020, Amrillah2023} and thermoelectric properties,\cite{Kim2017, Kumar2016} in different types of MXenes. 

It was revealed in the early studies that the bare \ce{Ti2C} monolayer acts as a conductor, whereas the structure with full saturation of oxygen atoms \ce{Ti2CO2} presents a semiconducting behavior. \cite{Khazaei2012, Amrillah2023}
In contrast, Champagne et al. \cite{Champagne2018} showed in their study that in the case of \ce{V2C} MXene, a metallic character can be observed for both the bare structure and the fully saturated one, which is present for all surface terminal groups. 
Furthermore, several studies point to the magnetic nature of pure carbide monolayers MXenes, while functionalization is done to remove the magnetism.\cite{Xie2013}
Zhang et al. \cite{Zhang2019} showed that the magnetic ground state of \ce{Mn2C} can be switched from antiferromagnetic (AFM) to ferromagnetic (FM) state by complete hydrogenation/oxygenation. 
Similarly, Bae \textit{et al.}\cite{Bae2021} has found the existence of magnetic states in other types of MXenes: \ce{V2CO2}, \ce{V2CF2}, and \ce{Mo2CF2}.
Several later studies also show that bare \ce{Ti2C} MXenes are AFM semiconductors \cite{Ketolainen2022, Peng} and surface termination of by the oxygen atoms modifies it from AFM to nonmagnetic (NM) phase.\cite{Bafekry, Ketolainen2022}
Likewise, He et al. \cite{He2016} revealed that the magnetic properties of \ce{Mn2CT2} (T = O, F, OH, Cl, and H) can be modulated by the surface functional group dependent on the electronegativity of the functional group. 
They showed that \ce{Mn2CT2} retains the FM ground state upon surface functionalization by groups with a formal charge of -1$e$ (T = F, Cl, and/or OH), whereas functional groups with a formal charge of -2$e$ (T = O) and +1$e$ (T = H) change the magnetic ground state to AFM. 

Moreover, a few of the defective MXenes become magnetic due to the presence of unpaired electrons in the spin-split d-orbitals. 
In 2D MXene exfoliation, etching with the MAX phase leads to the inevitable formation of defects, whose concentration can be controlled by adjusting the concentration of the etching chemicals during preparation.\cite{Wang2019} 
The defects can significantly affect the structural stability, electronic, and magnetic properties. 
Sang et al.\cite{Sang2016} experimentally showed that Ti vacancies affect the surface morphology and terminal groups and could be controlled by the etchant density. 
It was found that C vacancies increased the electronic conductivity\cite{Hu2017}and also induced the magnetism.\cite{Taoufik} 
Bandyopadhyay et al.\cite{Bandyopadhyay} theoretically demonstrated that point defects can be considered as a potential way to tune the magnetic and electronic properties of \ce{Ti2XT2} MXenes. 
Finally, Persson et al.\cite{Persson2019} experimentally achieved the removal of a significant number of terminal groups from the surface of \ce{Ti3C2T}$_x$. 
Thermal and environmental processing led to zero fluorine content and subsequently to a decreasing Ti:O ratio up to 3:0.6 (i.e., the surface coverage of ca. 30\%). 
Thus, it is evident that intrinsic point defects and terminal atom removal in MXene can emerge as a potential tool to modulate the properties of 2D layered MXenes toward promising device applications. 
All the aforementioned findings suggest that understanding the effect of functionalization as well as defect/desorption engineering on the unusual properties of MXenes materials is crucial to introducing their novel properties and extending their future applications. 
Although several theoretical studies have pointed to possible magnetic properties of MXenes and despite some of their parent MAX phases are magnetic (as \ce{Mn2GaC}, (Cr,Mn)$_2$AlC, or (Mo,Mn)$_2$GaC),\cite{Ingason2014, Mockute2014, Salikhov2017} these properties are largely unexplored experimentally. 
A breakthrough occurred in 2020 when experimental evidence of a magnetic transition in \ce{Cr2TiC2T}$_x$ was obtained.\cite{Hantanasirisakul2020}  Subsequently, magnetic behavior was also observed in the case of \ce{Ti3C2T2} doped with Nb, La, or Gd atoms.\cite{Fatheema2020, Iqbal2020, Rafiq2020} These findings pave the way for further studies of magnetism in this large family of 2D materials. The desired magnetic properties can potentially be achieved by tuning the transition metal, including C/N ratio, and/or surface terminations in MXenes. This multi-level control of magnetic properties is a unique advantage of MXenes compared to other existing 2D magnets.

Therefore, in this work, we have focused on the investigation of the pattern of adsorption as well as the desorption (or vacancy defects) of the terminal groups on the \ce{Ti2C} MXene surface. 
Afterward, we investigate the effect of terminal groups with different surface coverages and adsorption/vacancy patterns including linear, local, and partial patterns on the structural and magnetic properties of \ce{Ti2C} MXenes.
We have found that the pattern of adsorption/vacancy depends on the electronegativity of terminal groups. 
Interestingly, the magnetic properties of \ce{Ti2CT2} MXene depend not only on the character of the surface functional group, but also on the pattern of the vacancy (partial, local, and/or linear), and these properties can be chemically tuned for specific applications. 
In the context of these results, we were able to predict the dependence of the magnetic behavior on the surface coverage of functional groups on the \ce{Ti2C} MXene surface.

\section{Computational methods}
\label{sec:Methods}
The structural, and magnetic properties of all considered systems were performed within the context of spin-polarized density functional theory (DFT) implemented by Vienna \textit{ab initio} simulation package (VASP).\cite{Kresse1996a, Kresse1996b}
The exchange-correlation potential energies were described by the generalized gradient approximation (GGA) of Perdew–Burke–Ernzerhof (PBE) functional.\cite{Perdew97} 
In addition, we used the Hubbard-like U correction (GGA+U)\cite{Hubbard} because the simulations concerning correlation effects in transition metal systems were done. 
The U-value for Ti atoms was chosen as 4~eV based on a previous study\cite{Taoufik} in which this value was confirmed for \ce{Ti2CO2} MXene and is also a commonly used value for Ti oxides.\cite{Frey2019, Chakraborty} 
For validation, the meta-GGA strongly constrained and appropriately normed (SCAN) density functional\cite{Sun2015} was also used for the selected structures, especially in the case of \ce{Ti2CF2}, where the selected structures were also verified by PBE+U approach with the U-value of 2~eV. 
The calculations are performed with full structural optimization, where all atoms are relaxed in all directions using 5 $\times$ 5 $\times$ 1 supercell \ce{Ti50C25T}$_n$ (\textit{n} = 0 - 50, T = O and F (or exceptionally OH)). 
For simplicity, we will use the abbreviation $\mathrm{Ti_2C-n\cdot T}$ for the $n$ adsorbed T atoms in the rest of the text. 
The vacuum spacing was selected 15~Å between two adjacent layers to avoid interactions between the periodic images of slabs in the \textit{z}-direction. 
A plane-wave basis set (energy cutoff value of 400~eV) and the projector-augmented-wave (PAW) method were used.\cite{Kresse1999} 
A 2 $\times$ 2 $\times$ 1 and 4 $\times$ 4 $\times$ 1 grids for k-point sampling is used for structural optimization and electronic properties, respectively. 
Gaussian smearing of a maximum width of 0.005 eV was used. 
To ensure optimization and self-consistent convergence, the energy and force criteria are set to be 10$^{-7}$~eV and 10$^{-4}$~eV/Å, respectively. 
The optimized structures and spin density distribution were visualized using the VESTA code.\cite{Vesta} 
Relative energy $\Delta E$ is defined as the difference between the total energy of structures with higher energy and lower energy (assumed ground state). 
To investigate the stability of the adsorption of the atoms onto the surface of MXenes, the adsorption energies are calculated using 
\begin{equation}
E_\mathrm{Ad} = E_\mathrm{MXA}-(E_\mathrm{MX}+E_\mathrm{T}),
\label{eq:ad}
\end{equation}
where $E_\mathrm{MX}$ denotes the total energy of supercell of monolayer MXene ($\mathrm{Ti_2C-n\cdot T}$), $ E_\mathrm{MXA}$ represents the total energy of the MXene monolayer after the adsorption of the one additional terminal atom (T) onto the surface of MXene ($\mathrm{Ti_2C-(n+1)\cdot T}$), 
and $E_\mathrm{T}$ are total energies of isolated single atoms T (T = O, F or O+H) obtained from the same computational supercell.

\section{Results and Discussion}
Before the adsorption of terminal groups on \ce{Ti2C} MXene, we started by optimizing the position of the atoms and the lattice parameters of the bare \ce{Ti2C} MXenes. 
To verify the ground state of the bare structure, the optimization was performed for the corresponding possible nonmagnetic (NM), ferromagnetic (FM), and antiferromagnetic (AFM) configurations. 
The AFM configuration of the spin density on Ti atoms was selected based on the literature results. \cite{Peng, Shah}
The relaxed unit cell has a lattice constant of 3.07~{\AA}, and each C atom is covalently bonded to six neighboring Ti atoms with a bond length $d$(Ti--C) of~2.14 {\AA}. 
These results were in agreement with previous theoretical studies corresponding to a lattice constant value between 3.01 and 3.08~{{\AA}} and a bond length of approximately 2.10~{{\AA}}. \cite{Peng,Bafekry,Taoufik} 
Table \ref{tab:Tab1} shows the positive relative energy values for the FM and NM states, namely 0.06 and 0.28 eV, as compared to the AFM state, which is thus considered the ground state in agreement with the available literature.\cite{Ketolainen2022, Shah, Peng}  
\begin{table}[htbp]
\centering
\caption{Magnetic state (NM = nonmagnetic, FM = ferromagnetic, and AFM = antiferromagnetic), relative energy ($\Delta E$ in eV/unit cell), total magnetic moment ($m$ in $\mu_B$/unit cell) and magnetic moments for Ti atoms ($m_{Ti}$ in $\mu_B$) for the bare structure of \ce{Ti2C} MXene. The most stable geometry is highlighted in bold.} 
\begin{tabular}{lccccc}
\hline
Mag. St. & $\Delta E$ & $m$ &$m_{Ti}$ \\ 
\hline
NM &0.28& 0.00 & 0.00 \\
FM & 0.06 &  1.90 & 1.03 \\
\textbf{AFM} &\textbf{0.00}&  0.00 & 1.36\\ 
 \hline
\end{tabular} 
\label{tab:Tab1}
\end{table}

The self-consistent calculation after the structural optimization for the AFM configuration shows the partial magnetic moment on titanium atoms of value 1.36~$\mu_B$ and -1.36~$\mu_B$ (further only magnitude will be reported), 
while equal magnetic moments of each Ti atom of 1.03~$\mu_B$ is observed for FM solution (the small compensation of -0.16~$\mu_B$ on C atom lead to the total magnetic moment of 1.90~$\mu_B$ per unit cell; Table \ref{tab:Tab1})

On the other hand, in the case of fully covered surfaces, the \ce{Ti2CT2} MXenes (T = O, F, or OH), the magnetic states (AFM, FM) converge to a nonmagnetic solution during atomic relaxation and thus a nonmagnetic state is defined as the ground state. 
In the case of \ce{Ti2CO2} MXene, we observe the equilibrium bond length $d$(Ti--O) of 2.00~{\AA}, while in the case of \ce{Ti2CF2} and \ce{Ti2C(OH)2} MXene we observe bond lengths $d$(Ti--T) around 2.20~{\AA}. 
Moreover, after the adsorption of oxygen atoms (\ce{Ti2CO2}), Ti--C and Ti--Ti bond lengths are extended to 2.22 and 3.18~{\AA}, which indicates a strong interaction between oxygen and the surface of \ce{Ti2C} MXene. 
Last but not least, the stability of the structures was verified by the calculation of the cohesive energy according to Equation S1 in ESI: 
negative cohesive energy was observed for all terminal groups, indicating that all \ce{Ti2CT2} MXenes are stable and \ce{Ti2CO2} is most preferred (lowest cohesive energy, see Table S1). 
All the aforementioned results are consistent with previous studies presented.\cite{Bafekry, Shah} 

Several studies suggest that terminal groups, along with their vacancies, play a fundamental role in the properties of MXenes and may influence their magnetic, optical, and electronic behavior.\cite{Bafekry, Bandyopadhyay, Xiao, Taoufik, Venosova2023}
For this reason, in the next part of the study, we decided to investigate in more detail the different adsorption patterns of the terminal groups as well as their vacancy/desorption patterns. 
In line with this aim, the effect of surface coverage and adsorption/desorption pattern on the magnetic behavior of \ce{Ti2CT2} MXenes was investigated.

\subsection{Adsorption patterns}
\begin{figure}[h]
 \centering
 \includegraphics[width=8.5cm]{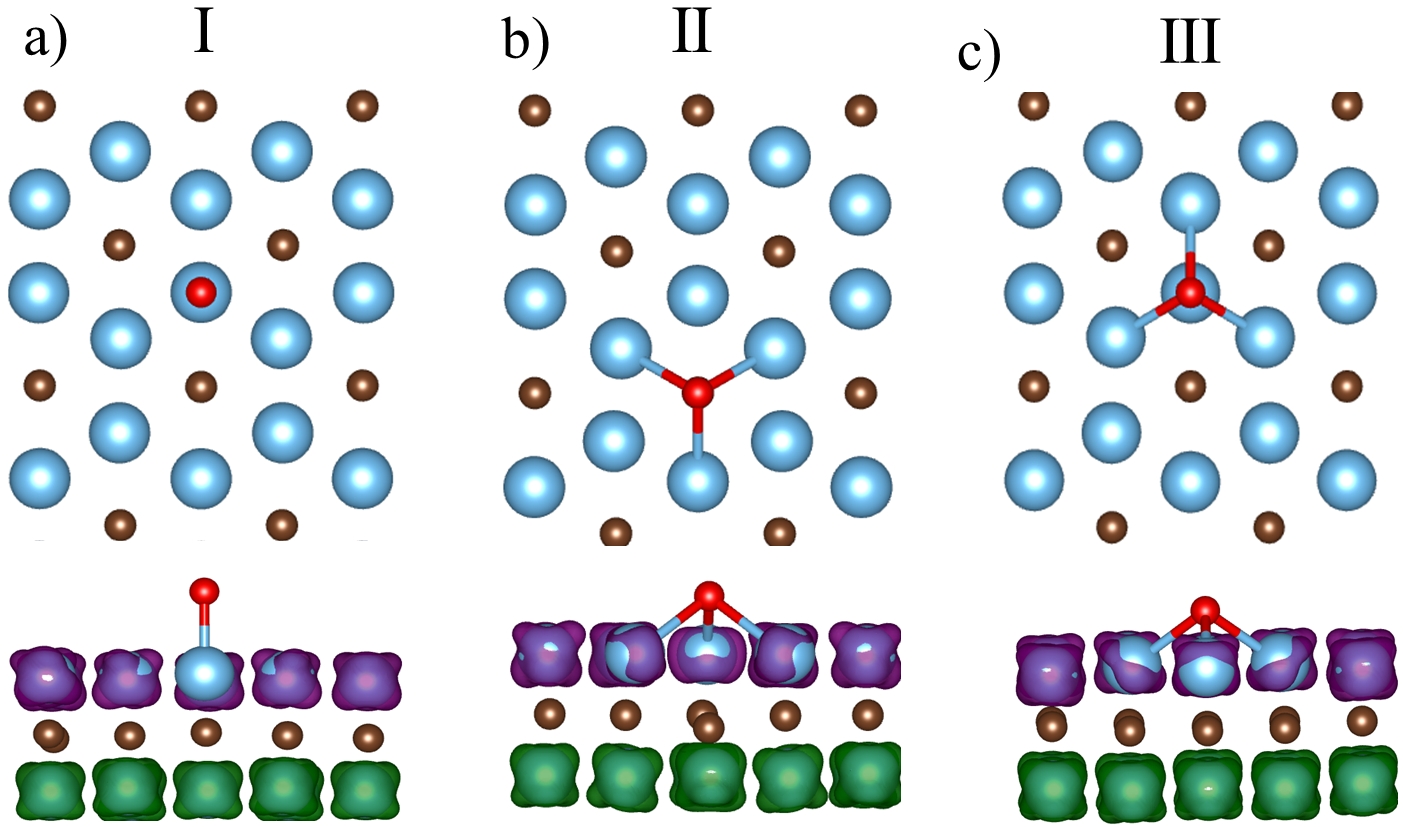}
 \caption{Top and side view of the bare \ce{Ti2C} MXene after adsorption of one oxygen atom in 5 x 5 supercells on a) metal site, b) carbon site, and c) hollow site. Spin density distribution 
 was added to the side view -- green and purple colors represent positive (spin up) and negative (spin down) spin densities, respectively, and isosurface values are 0.02 e/\AA$^{3}$.
 Titanium, carbon, and functional oxygen atoms are represented in blue, brown, and red, respectively.}
 \label{fig:1Atom}
\end{figure}

In the following, we systematically investigate the structural and magnetic properties of $\mathrm{Ti_2C-n\cdot T}$ monolayers (using supercell \ce{Ti50C25T}$_n$ models, T = O, F). Different surface coverage \textit{c} of terminal atoms T were considered (where \textit{c} = 2 – 100~\% for \textit{n} = 1 – 50 atoms, respectively) and various adsorption patterns including partial, linear, and local motifs were investigated. 
First, the most probable adsorption position of the single functional atoms was investigated (\textit{c} = 2\%, \ce{Ti50C25T1} supercell). 
In \ce{Ti2C} MXene, there are three possible high-symmetric adsorption sites (models I, II, and III, see Figure \ref{fig:1Atom}). 
In the first model (I), the functional group is directly located on top of the titanium atom (metal site, see Figure \ref{fig:1Atom}a). 
In the second model (II), the functional group is located above the carbon atom between the three neighboring Ti atoms (carbon site, see Figure \ref{fig:1Atom}b). 
Similarly, in the third model (III), the functional group is located above the bottom Ti atom (hollow site, see Figure \ref{fig:1Atom}c). 
In the case of -OH terminations, the H atom is placed on top of an oxygen atom, making the functional group termination equivalent to mono-atom termination. 
First, the relative stability of the optimized MXenes with one terminal atom was verified for all the above configurations and three different terminal atoms/groups (O, F, and OH). 
In general, our results are consistent with previous studies.\cite{Shah,Zhang2018} 
The relative energies ($\Delta E$) of the studied MXenes are listed in Table \ref{tab:Tab2}. 
\begin{table}[htbp]
\caption{Magnetic states (NM = nonmagnetic, FM = ferromagnetic and AFM = antiferromagnetic), relative energies ($\Delta E$ in eV per 5$\times$5 supercell), and adsorption energy ($E_\mathrm{Ad}$ in eV/atom) for the $\mathrm{Ti_2C-1\cdot T}$ (T = O, F and OH).} 
\setlength{\tabcolsep}{2.3pt} 
\renewcommand{\arraystretch}{1.1} 
\begin{tabular}{lccc|ccc|ccc}
\hline
 &\multicolumn{9}{c}{Relative energies $\Delta E$ (in eV/supercell)} \\ \hline
& \multicolumn{3}{c|}{O} & \multicolumn{3}{c|}{F} & \multicolumn{3}{c}{OH} \\
 & I & II & \multicolumn{1}{c|}{III} & I & II & \multicolumn{1}{c|}{III} & I & II & III \\ \hline
AFM* & 2.31 & 0.39 & \multicolumn{1}{c|}{0.00} & 1.07 & 0.17 & \multicolumn{1}{c|}{0.00} & 1.34 & 0.12 & 0.00 \\
FM & 3.75 & 1.88 & \multicolumn{1}{c|}{1.18} & 2.27 & 1.66 & \multicolumn{1}{c|}{1.33} & 2.63 & 1.64 & 1.28 \\
NM & 9.24 & 7.06 & \multicolumn{1}{c|}{6.71} & 7.90 & 7.26 & \multicolumn{1}{c|}{6.91} & 8.08 & 7.06 & 6.74 \\ \hline
\multicolumn{10}{c}{Adsorption energy of a single atom $E_\mathrm{Ad}$ (in eV/supercell)} \\ \hline
 & \multicolumn{3}{c|}{O} & \multicolumn{3}{c|}{F} & \multicolumn{3}{c}{OH} \\
 & I & II & \multicolumn{1}{c|}{III} & I & II & \multicolumn{1}{c|}{III} & I & II & III \\ \hline
AFM* & -6.03 & -7.94 & \textbf{-8.33} & -5.36 & -6.26 & -6.43 & -4.10 & -5.33 & -5.44 \\
FM & -4.59 & -6.60 & -7.15 & -4.16 & -4.76 &-5.10 & -2.81 & -3.81 & -4.12 \\
NM & 0.90 & -1.28 & -1.63 & 1.47 & 0.84 & 0.48 & 2.64 & 1.62 & 1.30 \\ \hline
\end{tabular}
$^*$ AFM represents the initial magnetic state before atomic relaxation
\label{tab:Tab2}
\end{table}
In the three geometries, the structures with geometry III have the lowest total energy for all studied terminal groups, demonstrating that geometry III is energetically more favorable than geometries I and II. 
Therefore, position III can be assumed as the most favorable for the formation of strong chemical bonds in MXenes. 
Subsequently, according to Equation \ref{eq:ad}, the adsorption energy was calculated to determine the stability and preference of the terminal groups. 
The most stable terminal group is oxygen with the lowest $E_\mathrm{Ad}$ (-8.33~eV), followed by fluorine ($E_\mathrm{Ad}$ = -6.4~eV), and the OH group has the highest energy ($E_\mathrm{Ad}$ = -5.44~eV), making it the least preferred terminal group. 
The lowest $E_\mathrm{Ad}$ values for O-functionalized \ce{Ti2C} may be due to the stronger interaction between the oxygen and titanium atom, which results from the shorter bond length of Ti--O (2.09~{\AA}) than those of Ti--F (2.16~{\AA}) and Ti--OH (2.18~{\AA}). 
In Table \ref{tab:Tab2}, positive values of the adsorption energy are observed in the case of non-magnetic states, indicating that the non-magnetic states are not stable. 
For all adsorption sites as well as for all terminal atoms, the ground state is observed as the one growing from the antiferromagnetic solution (see Table \ref{tab:Tab2}). 
However, a non-zero net magnetic moment (from 0.5 to 1.75~$\mu_B$ per supercell; see Figure S1 for spin densities) is observed for all considered configurations (adsorption sites and terminal groups). 
The spin density is redistributed after adsorption on the surface due to charge transfer between the terminal atom and the adjacent titanium atoms (e.g. charge loss on Ti atoms 0.34$e$ in the case of adsorbed oxygen). 
This distortion leads to the transition from the antiferromagnetic state with zero total magnetic moments to the not fully compensated antiferromagnetism, i.e., to the ferrimagnetic (FiM) state with a total magnetic moment of 1.75~$\mu_B$, 1~$\mu_B$, and 1~$\mu_B$ per supercell (configuration III as the ground state, see also Figure S1) for $\mathrm{Ti_2C-O}$, $\mathrm{Ti_2C-F}$, and $\mathrm{Ti_2C-OH}$, respectively. 
This is also visually supported by Figure \ref{fig:1Atom} -- the overall antiferromagnetic character remains preserved and there is a local weakening of spin density (spin density hole) located on three adsorbate-neighboring Ti atoms. 
These results also indicate that the magnetic properties directly depend on the formal oxidation state of the terminal group. 

Significant differences can also be observed in the case of the electronic structure. Figures S13 - S14 show the projected density of states (PDOS) of selected monolayers. 
We can notice that in the case of adsorption of one oxygen atom, the AFM semiconductor turns into a half-metal: the monolayer ${\mathrm{Ti_2C-1}\cdot O}$ exhibits a semiconducting character for spin-up and a metallic behavior for spin-down. 
On the contrary, the semiconducting character is preserved in the case of $\mathrm{Ti_2C-1\cdot F}$.
Similarly, He et al. \cite{He2016} have demonstrated the possibility of tuning the magnetic properties as a function of the electronegativity of the functional groups (a group bearing a formal charge of -1$e$ (F, OH) versus a group bearing a formal charge of -2$e$ (O)). 
The obtained results as well as the literature\cite{Ibragimova, Shah, Hu2018} indicate that the -OH terminal group is the least preferred and at the same time, a similar behavior as for terminal atoms -F is assumed.\cite{Shah, He2016} 
For this reason, in the next part of the study we decided to investigate and compare only the differences between the -O and -F terminal atoms in detail, and the -OH terminal group is not discussed further. 
\begin{figure}[htbp]
\centering
\includegraphics[width=9cm]{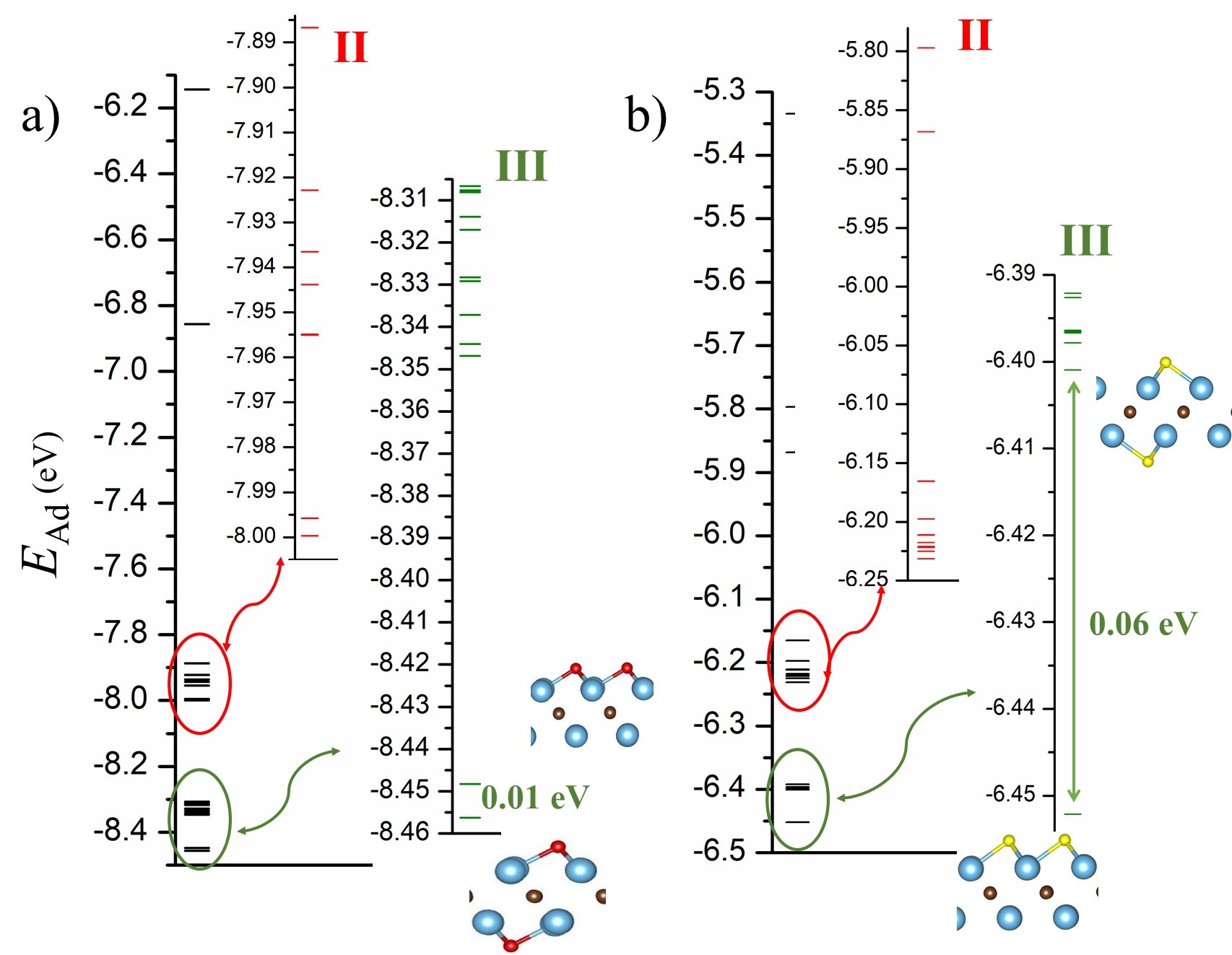}
\caption{Energy diagram for adsorption of the second (a) O atom, and (b) F atom on the various adsorption sites. The adsorption energy $E_\mathrm{Ad}$ in position II [III] is represented by the magnified energy diagrams on the left [right] side (red [green] color). The inset figures show the most preferred structures with the position of the second atom (labeled Ti1 and OTi1 according to the notation in the ESI Figure S2). Blue, brown, red, and yellow colors refer to titanium, carbon, oxygen, and fluorine atoms, respectively. The most preferred initial magnetic ground state was considered (AFM; see Tables S2 and S3 in ESI).}
\label{fig:2AtomsCE}
\end{figure}

\begin{figure*}[h]
\centering
 \begin{subfigure}[b]{\textwidth}
 \centering
 \caption{\ce{Ti2C} + \textit{n}$\cdot$O (\textit{n} = 1 -- 20)}
 \includegraphics[width=15cm]{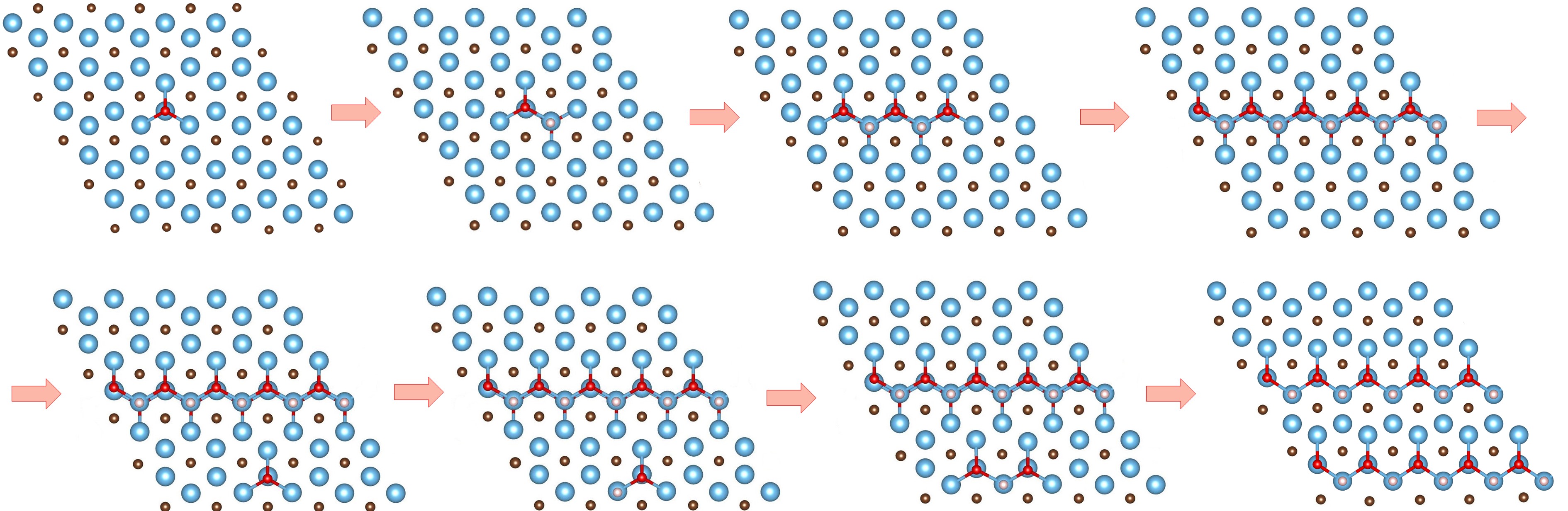}
 \label{fgr:PattOa}
 \end{subfigure}
 \begin{subfigure}[b]{\textwidth}
 \centering
 \caption{\ce{Ti2C} + \textit{n}$\cdot$F (\textit{n} = 1 -- 15)} 
 \includegraphics[width=15cm]{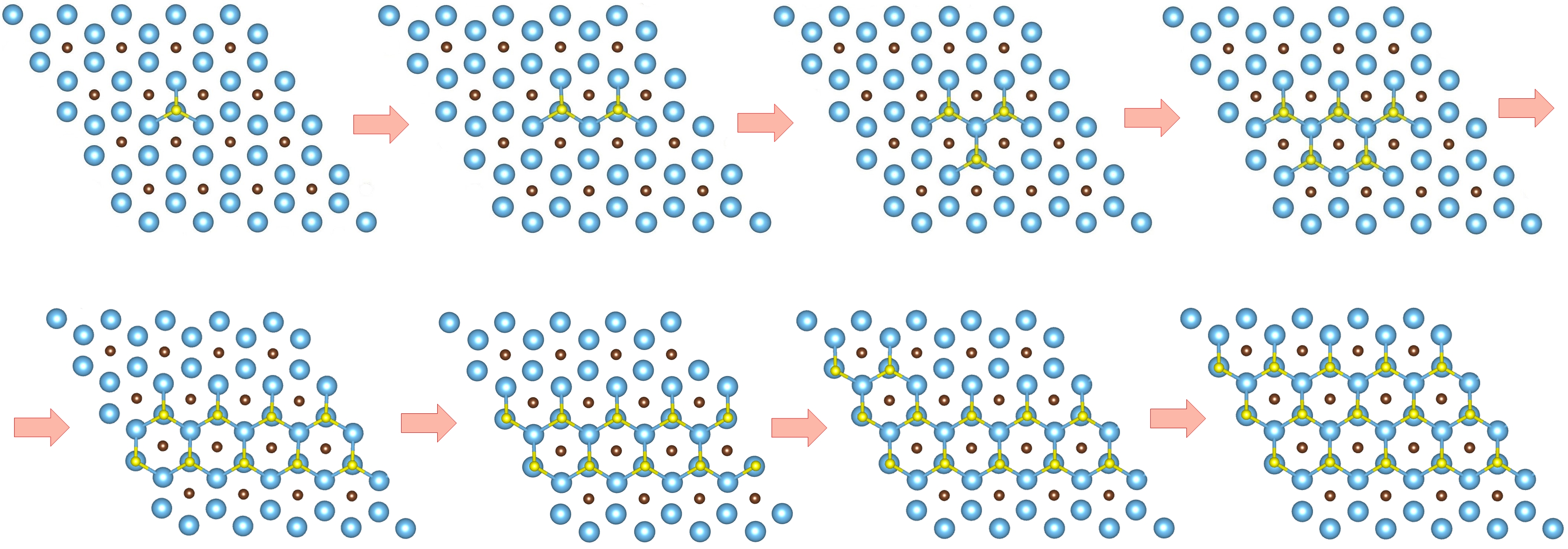}
 \label{fgr:PattOb}
 \end{subfigure}
 \caption{Adsorption pattern of the \ce{Ti2C} MXene surface during (a) oxidation and (b) fluorination up to 20 termination atoms (representing a surface coverage from 2 to 40\%). Blue, brown, red, and, yellow colors refer to titanium, carbon, oxygen, and fluorine atoms, respectively.}
 \label{fgr:PatO}
\end{figure*}
In the following procedure, a second atom in different positions was added to the terminal atom in position III (the most suitable position) to find the most likely adsorption positions for both (the individual positions and their labels for \ce{Ti50C25T2} supercell are shown in Figure S2). 
To verify the preference for position III, all three possible initial positions I-III were again considered. 
Moreover, the distance of the added atom from the first atom was gradually increased as well as the opposite side was tested (see Figure S2). 
This results in 23 configurations of one termination type with corresponding energies visualized in Figure \ref{fig:2AtomsCE} (relative energies and magnetic moments are summarized in Tables S2 and S3 for $\mathrm{Ti_2C-2\cdot O}$ and $\mathrm{Ti_2C-2\cdot F}$ configurations, respectively). 
Importantly, it should be noted that in most cases, when the second atom was placed in position I, the atom moved during relaxation and the final geometry represented the position of the second atom in position II or III (see Figures S3 - S4 and Tables S2 - S3 for $\mathrm{Ti_2C-2\cdot O}$ structures). 
This demonstrates that the position of the second atom in position I is again highly unstable and unlikely. 
The values of adsorption energy presented in Figure \ref{fig:2AtomsCE} are all negative, indicating that all structures for two terminal atoms in positions II and III are stable. 
The hollow site (III) of the second atom is preferred over the carbon site (II), as documented in Figure \ref{fig:2AtomsCE} (and Tables S2-S3) by splitting of lowest energies into two groups (magnified red and green energy diagrams in Figure \ref{fig:2AtomsCE}). 
Figure \ref{fig:2AtomsCE} also shows the energy proximity of the two lowest energy structures with different positions of the second atom (difference of only 0.01 and 0.06~eV for $\mathrm{Ti_2C-2\cdot O}$ and $\mathrm{Ti_2C-2\cdot F}$, respectively). 
\begin{table}[h]
\centering
 \caption{Magnetic states (NM = nonmagnetic, FM = ferromagnetic and AFM = antiferromagnetic), relative energies ($\Delta E$ in eV/supercell 5 x 5) of \ce{Ti2C} - T with different surface coverages of adsorbed terminal atoms (oxygen and fluorine).} 
\begin{tabular}{cccccccc}
\hline
\multicolumn{8}{c}{10\% coverage of terminal atoms} \\ \hline
\multicolumn{4}{c|}{\ce{Ti2C} - 5O} & \multicolumn{4}{c}{\ce{Ti2C} - 5F} \\ \hline
 & AFM* & FM & \multicolumn{1}{c|}{NM} & & AFM* & FM & NM \\ \hline
A & 0.39 & 1.20 & \multicolumn{1}{c|}{5.64} & A & 0.15 & 0.94 & 5.68 \\
B & 0.00 & 0.78 & \multicolumn{1}{c|}{5.09} & B & 0.00 & 0.86 & 5.78 \\
C & 0.25 & 1.04 & \multicolumn{1}{c|}{5.30} & C & 0.04 & 0.91 & 5.82 \\
D & 0.01 & 0.80 & \multicolumn{1}{c|}{5.12} & D & 0.24 & 1.05 & 5.76 \\
E & 0.13 & 0.79 & \multicolumn{1}{c|}{5.06} & E & 0.30 & 1.08 & 5.78 \\ \hline
\multicolumn{8}{c}{40\% coverage of terminal atoms} \\ \hline
\multicolumn{4}{c|}{\ce{Ti2C} - 20O} & \multicolumn{4}{c}{\ce{Ti2C} - 20F} \\ \hline
 & AFM* & FM & \multicolumn{1}{c|}{NM} & & AFM* & FM & NM \\ \hline
A & 3.06 & 2.99 & \multicolumn{1}{c|}{5.32} & A & 1.01 & 1.39 & 3.69 \\
B & 0.00 & 0.03 & \multicolumn{1}{c|}{0.86} & B & 1.57 & 1.39 & 2.67 \\
C & 5.14 & 5.25 & \multicolumn{1}{c|}{8.43} & C & 0.20 & 0.00 & 3.12 \\
D & 1.78 & 1.75 & \multicolumn{1}{c|}{2.95} & D & 1.70 & 1.57 & 3.60 \\
 & & & \multicolumn{1}{c|}{} & E & 0.18 & 0.11 & \\ \hline
\multicolumn{8}{l}{*the initial magnetic state before optimization}
\end{tabular}
\label{tab:Tab3}
\end{table}
We note that the energy difference between the preferred hollow (III) site configuration and lowest carbon (II) site is approximately 0.50~eV and 0.25~eV for $\mathrm{Ti_2C-2\cdot O}$ and $\mathrm{Ti_2C-2\cdot F}$, respectively. 
Based on these results, only adsorption to hollow (III) positions was considered in the following study with different patterns including partial, linear, and local patterns.
Concerning magnetism, one-side adsorption leads to the summing of the local magnetic effect and an increase of the total magnetic moment (in the range from 2.0~$\mu_B$ to 6~$\mu_B$ per supercell; Tables S2 and S3), while both-side adsorption compensates local magnetic moments to the antiferromagnetic arrangement (in fact, two spin density holes of Figure \ref{fig:1Atom} cancel one other). 

To determine the adsorption patterns on the surface of \ce{Ti2C} MXene, the atoms were gradually added in different patterns including linear, local, and partial patterns. 
Several adsorption positions (from 5 to 10 patterns per surface coverage \textit{c}) were tested during the gradual addition of atoms, i.e., increasing coverage (the obtained patterns are shown in Figures S5 – S12 together with associated relative energies and magnetic moments in Tables S4, and S5).
The results show that the range of energy differences between the patterns (assuming the comparison of only the ground magnetic state (AFM and/or FM) for individual structures) increase with increasing surface coverage up to 40\%. 
Table \ref{tab:Tab3} shows an example that in the case of \textit{c} = 10~\% of oxygen atoms, maximum energy differences between the patterns are at most 0.39 eV, while in the case of \textit{c} = 40~\%, a maximum energy difference is observed of up to 5.14 eV. 
In the case of surface coverage with fluorine atoms, an increase in maximum energy differences from 0.30~eV (\textit{c} = 10~\%) to 1.57~eV (\textit{c} = 40~\%) is observed too. 
The increasing energy differences illustrate that in the case of higher coverages, the probability of an adsorption pattern can be assumed more easily. 

\begin{figure}[h]
\centering
 \includegraphics[height=6.5cm]{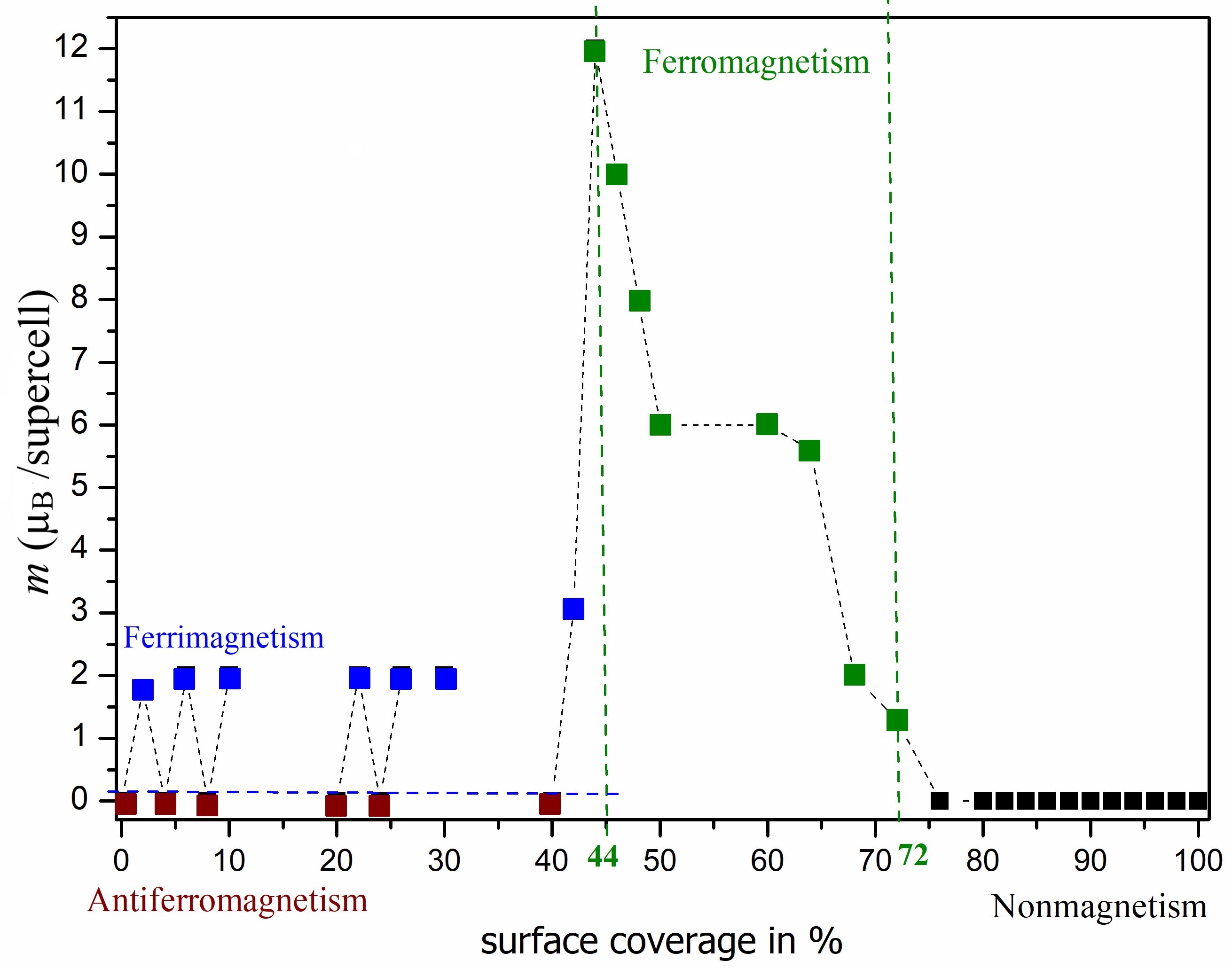}
 \caption{Dependence of magnetic character on surface occupancy with a total magnetic moment for the supercell of \ce{Ti2C} - \textit{n}O MXenes (where \textit{c} = 2 – 100 \% for \textit{n} = 1 – 50 atoms). Dashed lines are a guide to the eye.}
 \label{fgr:MMO}
\end{figure}

\begin{figure}[h]
\centering
 \includegraphics[height=6.5cm]{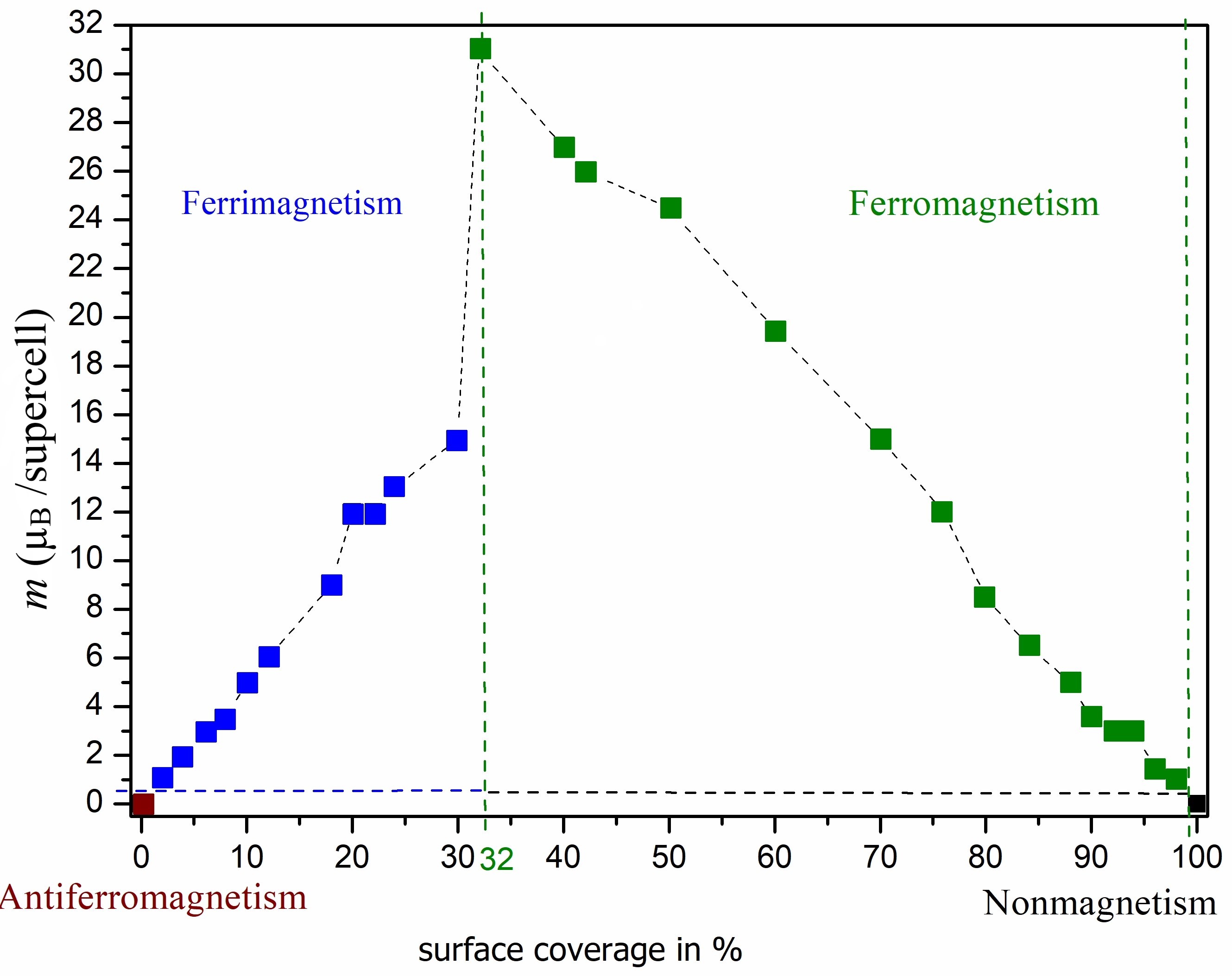}
 \caption{Dependence of magnetic character on surface occupancy with a total magnetic moment for the supercell of \ce{Ti2C} - \textit{n}F (where \textit{c} = 2 – 100 \% for \textit{n} = 1 – 50 atoms). Dashed lines are a guide to the eye.}
 \label{fgr:MMF}
\end{figure}

Based on the results obtained, we were able to predict the likely pattern of adsorption of atoms on the surface (see Figure \ref{fgr:PatO}). 
Figure \ref{fgr:PattOa} shows that a zigzag line (double-faced) pattern is preferred in the case of oxygen adsorption. 
Furthermore, it is interesting to observe that when one row is filled, no adsorption occurs immediately next to it, but a gap is formed between the occupation rows. 
However, this is also influenced by our finite size model (supercell \ce{Ti50C25T}$_n$) and should be considered with caution. 
Figure \ref{fgr:PattOb} clearly shows that a different pattern is observed in the case of the adsorption of fluorine atoms on surfaces bearing a -1$e$ formal charge. 
In this case, the adsorption of atoms only from one side is preferred, i.e., the local flake pattern is observed as the most stable. 
Significant differences can also be observed in the case of magnetic behavior (see Figures \ref{fgr:MMO} and \ref{fgr:MMF}). 
In the case of O-termination, magnetism depends on the number of atoms as shown in Figure \ref{fgr:MMO}. 
While in the case of even numbers of oxygens on the surface, the ground state is antiferromagnetic, the ferrimagnetic state is there for structures with an odd number of oxygen atoms on the surface -- but this is still basically antiferromagnetic arrangement not compensated just in one site (an analogy to $\mathrm{Ti_2C-2\cdot O}$ vs. $\mathrm{Ti_2C-O}$). 
We note that the energy differences between the AFM/FiM state and FM state decrease with increasing surface coverage from 1.03~eV (2\%) to 0.03~eV (40\%; see also $\Delta E$ in Table 3 and S4). 
At 44\% surface oxygen coverage, we observe the sudden change in the ground state from AFM/FiM to FM with a total magnetic moment of 12 $\mu_B$ -- the mechanism of compensation of local spin density holes stopped. 
With increasing surface coverage the total magnetic moment of ferromagnetic structures decreases to zero and for \textit{c} > 72\% a change from a ferromagnetic ground state to a nonmagnetic one is observed. 
One-side "flake" adsorption of fluorine atoms (Figure \ref{fgr:PattOb}) has different consequences on magnetism in $\mathrm{Ti_2C-n\cdot F}$. 
There is a linear increase in the magnetic moment (from 0 to 12~$\mu_B$) due to the summing of the local magnetic effect -- local spin density holes are not compensated, but create an island with zero magnetic density over titanium atoms under adsorbed fluorine atoms. 
This behavior stops if the surface coverage is greater than 30\%. 
At increased surface coverage above 32\%, a change to a ferromagnetic state is observed until the surface is completely covered (to be discussed in Section 3.2). 
Moreover, in the case of $\mathrm{Ti_2C-n\cdot F}$, the magnetic moment (31~$\mu_B$) is almost twice compared to $\mathrm{Ti_2C-n\cdot O}$. 
Thus, the results suggest that the magnetism depends on both the coverage and the type of terminal atoms. 
In addition, it is interesting to observe that if only one Ti layer of MXenes is covered, the opposite (uncovered) Ti layer is not affected. 

The dependence of the electronic properties on the number of atoms as well as on the type of terminal atoms can also be observed (see Figures S13 -- S14 in ESI). 
In the case of an even number of O atoms, a semiconducting character is observed. Conversely, in the case of an odd number of O atoms, a half-metal or conducting character is observed. 
With increasing surface atoms, the band gap decreases, with a significant change occurring at 20\% coverage, where a change to a conducting character is achieved (though the AFM is preserved). 
The conducting character is preserved with the increase in the number of atoms, resulting in a change of magnetic alignment (change to FM). 
Similarly, in the case of F atoms, a decrease in the band gap can be observed with an increasing number of atoms on the surface. This is followed by a change from semiconductor to conductor character together with a change in magnetic character (ferri- to ferromagnetic). 
The conducting character, as well as the FM, is retained until the surface is completely covered (discussed in the next section).


\begin{figure*}[htb]
 \centering
 \begin{subfigure}[b]{\textwidth}
 \centering
 \caption{\ce{Ti2CO} -- \textit{n}V$_O$ (\textit{n} = 1 -- 20)}
 \includegraphics[width=15cm]{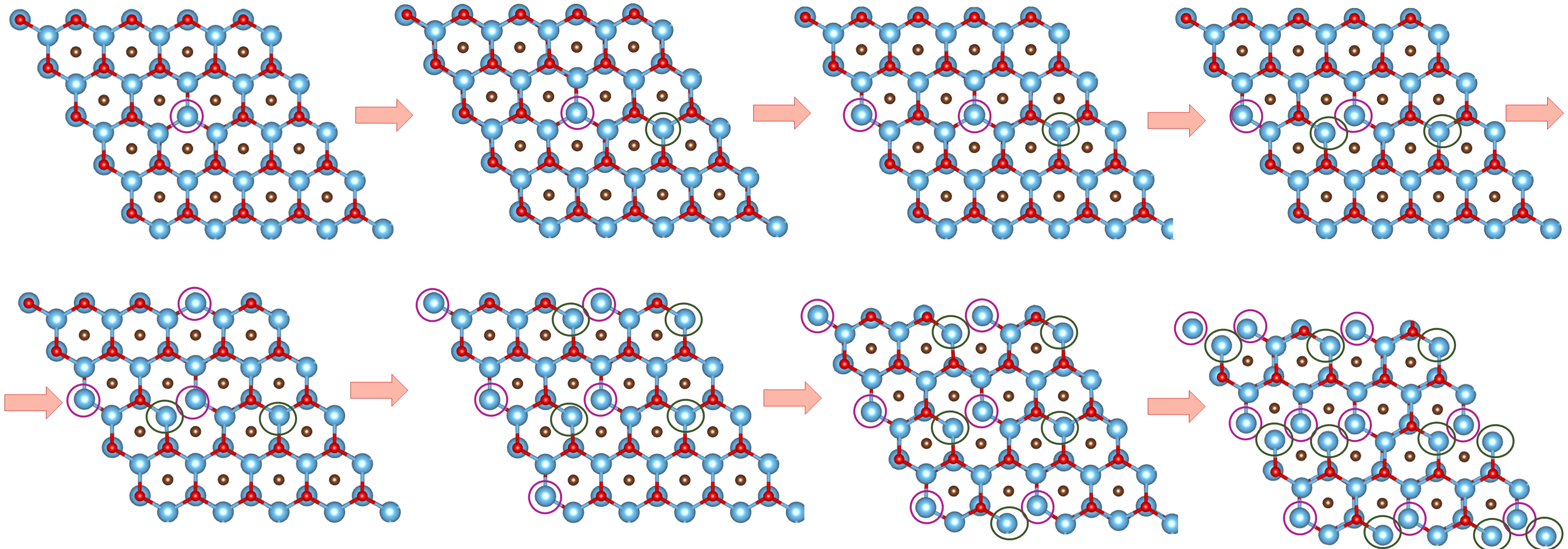}
 \label{fgr:PattVacOa}
 \end{subfigure}
 \begin{subfigure}[b]{\textwidth}
 \centering
 \caption{\ce{Ti2CF} -- \textit{n}V$_F$ (\textit{n} = 1 -- 15)}
 \includegraphics[width=15cm]{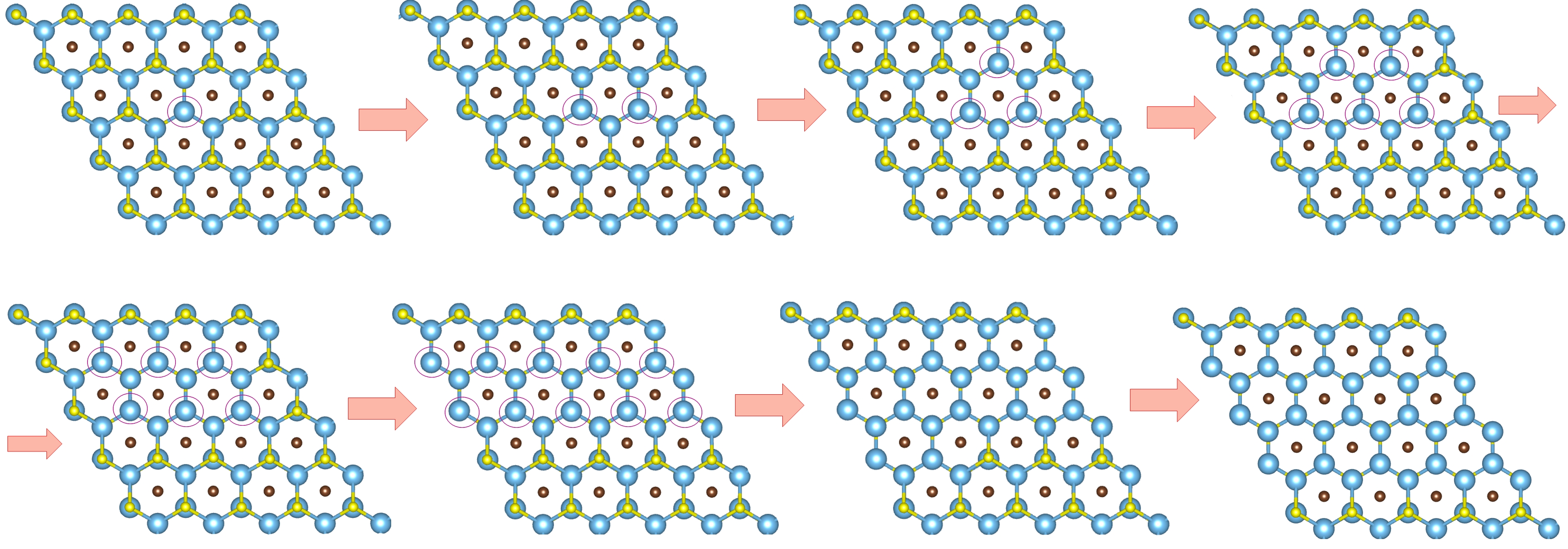}
 \label{fgr:PattVacOb}
 \end{subfigure}
 \caption{Vacancy pattern on the surface of MXene \ce{Ti2CT2} (where T = O, F) up to 20 vacancies defect (representing a vacancy concentration from 2 - 40 \%). The pink circle represents the defect of the atom from the top side and the green circle represents the defect from the opposite side. Titanium, carbon, functional oxygen, and fluorine atoms represented blue, brown, red, and yellow respectively.
 }
 \label{fgr:PattVacO}
\end{figure*}
\subsection{Vacancy defect patterns}
\begin{figure}[h]
\centering
 \includegraphics[width=9cm]{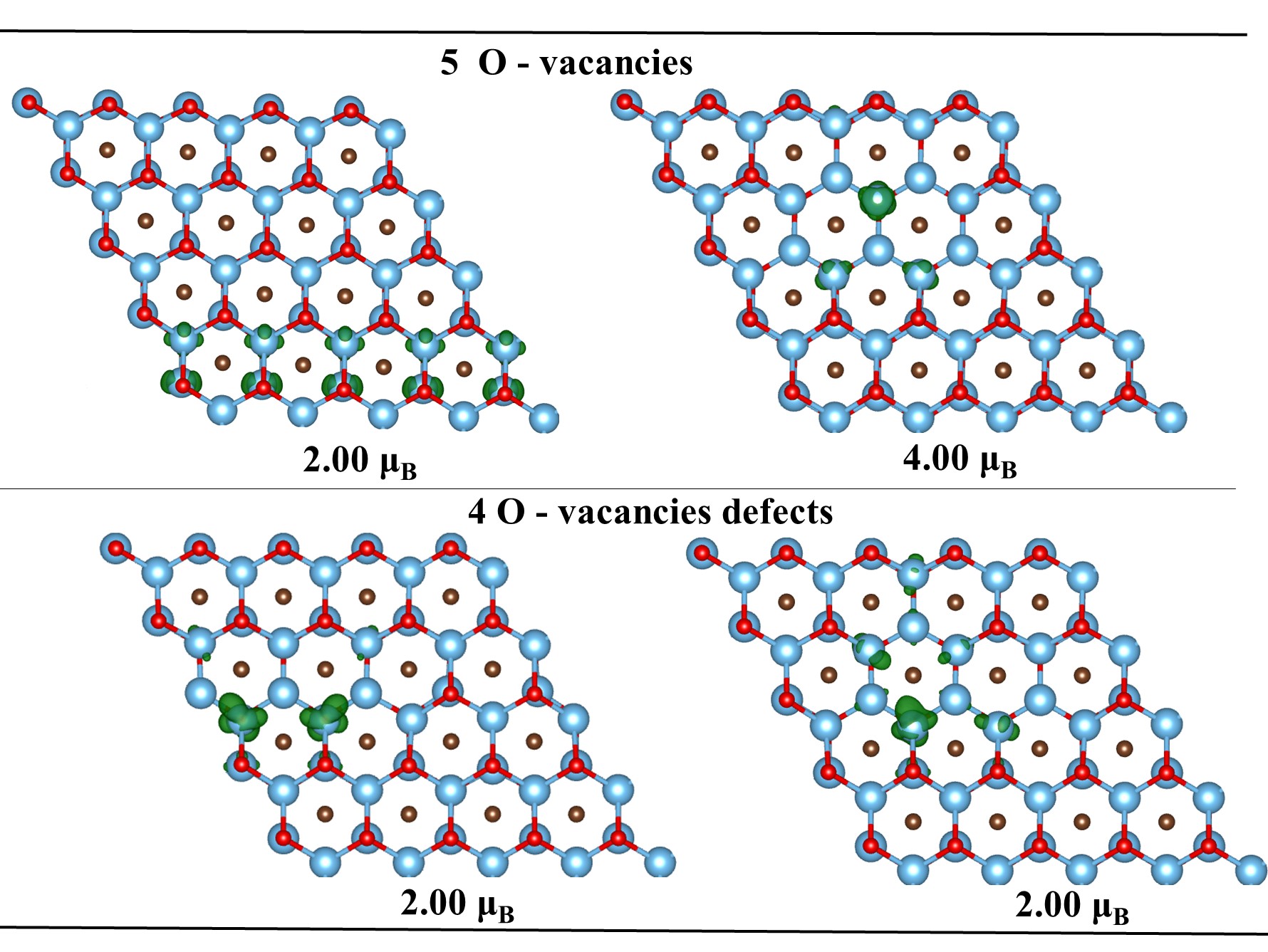}
 \caption{Spin density distributions for linear O-vacancies (left) and local O-vacancies (right). The isosurface values are 0.02 e/{\AA}$^{3}$ (purple color represents negative spin density). Titanium, carbon, and functional oxygen atoms represented blue, brown, and red, respectively.}
 \label{fgr:Spin45}
\end{figure}

In the next part of the study, we focused on a more detailed investigation of the influence of vacancy concentration and their pattern on the structural and magnetic properties of \ce{Ti2CT2} (T = O and F) monolayers. 
To investigate T atom vacancies, T atoms were removed to produce different concentrations of vacancies for different defect patterns including partial, linear, and local ones. 
To find suitable patterns for the vacancy defects, 5 –- 10 different patterns for each vacancy concentration were examined sequentially, following the same procedure as for adsorption (see Tables S6 and S7). 
Based on this, the most likely vacancy patterns were generated for \ce{Ti2CO2} and \ce{Ti2CF2} (see Figure \ref{fgr:PattVacO}). 
In the case of the O-vacancy defect, similarly to adsorption, we can observe a predominant zigzag orientation, however with a partial pattern where no alignment of the defect to a given single line is observed (see Figure \ref{fgr:PattVacOa}).
Compared to adsorption, it can be noticed that the removal of one oxygen atom does not cause a change in the magnetic properties of the material (NM character is retained). This is because, in the case of a single O vacancy, two electrons are lost, which do not cause a significant disruption of the structure and no local magnetic moment is generated.
The partial defect pattern leads to the formation of smaller defects further apart, with the loss of opposite-oriented oxygen, where the loss of charge is compensated (similar to adsorption) and no magnetic moment is generated. 
Moreover, based on the partial density of states (PDOS), we can also observe that there is no change in the semiconducting nature of the material (see Figure S15).
Subsequently, we can observe changes in the surrounding Fermi level and with increasing O-vacancy concentration a change from semiconducting to conducting character (at approximately 20\% of the defect surface). 
We can see that DOS near the Fermi surface comes mainly from the Ti-d and O-p orbitals. 
The reason may be that the introduction of oxygen vacancies leads to more charge transfer for neighboring O and Ti atoms. 
As a consequence, we observe that the non-magnetic state is maintained up to a defect vacancy concentration of 28\% (i.e., 72\% of the surface coverage, see Figure \ref{fgr:MMO}), where only vacancies consisting of at most two atoms are still observed. 
As the defect is enlarged further, more atoms (at least three) are involved in the vacancy resulting in the creation of a weak local moment on the neighboring Ti atoms. The subsequent increase in vacancy also causes an increase in the total magnetic moment with ferromagnetic spin ordering, due to the increasing spin polarization at the vacancy sites. 
The presence of spin polarization in the case of a defect in which a higher number of neighboring oxygen atoms is removed is confirmed by a change in the defect pattern. 
In the case of linear and local defect vacancies, spin polarization occurs where we observe a weak total magnetic moment of 2~$\mu_B$ already at 8\% vacancy concentration. 
In Figure \ref{fgr:Spin45} we can observe the emergence of a weak local magnetic moment in the case where we have vacancies localized close to each other (linear and local pattern). 
In the case of the local defect, we also observe a change from a semiconducting character to a half-metal character. Where similarly as in the case of adsorption we observe a semiconducting character for spin-up and a metal character for spin-down (see Figure S16).
This leads us to conclude that in the case of local/linear vacancies, even at low vacancy concentrations, it is possible to induce a magnetic moment and ferromagnetic character. 
Moreover, local O-vacancy defects can lead to the formation of half-metal characters. 
Similar characteristics have been studied in the case of other types of metal carbides, but also in the case of Ti\cite{Bafekry}.

In the case of F-vacancy, the defect vacancy pattern is observed in the same manner as in the adsorption case (i.e., local surface flake defect, see Figure \ref{fgr:PattVacOb}). 
The presence of spin polarization and total magnetic moment are already observed at the formation of a single vacancy. 
This is likely because the removal of only one fluoride leads to a loss of one electron to the surface, resulting in weak spin polarization around the vacancy. 
With increasing vacancy defect concentration, an increase in the total magnetic moment can also be observed with ferromagnetic spin alignment. 
It can also be observed that there is no change in the conductor character in the case of vacancy F (see Figure S15). The conducting character with FM is probably the most stable state for MXenes when the surface is terminated with F atoms. Thus, these results suggest that the presence of F atoms on the surface of MXenes may very likely contribute to the conducting character even in the case of end-group mixing. This may also lead to a higher possibility of magnetic moment generation with the FM arrangement. 
These results are also in agreement with other works.\cite{Sakhraoui2024, Bafekry}

Nevertheless, it should be noted that in the case of F-vacancies, the spin density distribution was incorrect when using the PBE+U(4eV) functional for the lowest F-vacancy concentrations -- artificial density distribution over the entire material in a ferrimagnetic spin alignment appeared (Figure S17). 
We therefore recalculated problematic configurations using PBE+U(2eV) and SCAN density functionals, which provided standard spin density localized at the vacancy sites with a ferromagnetic alignment. 
The incorrect description of the spin density distribution from PBE-U(4eV) did not affect the determination of the preferred pattern of vacancies, which was observed equally for all density functionals. 
Control calculations were also performed for adsorption as well as for $\mathrm{Ti_2C-n\cdot O}$, where the choice of functional did not affect the results. 
It can also be observed that the spin polarization is localized only on the Ti atoms where the vacancy occurs and that the opposite Ti layer is not affected.

Thus, the results show that magnetism is dependent on the electronegativity of the functional groups and the vacancy defect pattern. 
However, it should be noted that the cohesive energies show that fully covered MXenes are the most stable for all terminal groups and the cohesive energy decreases with an increasing number of terminal groups (see Figure S18). 
Nevertheless, these results suggest that the magnetic behavior can be tuned by modeling the surface according to the type of functional atoms, their concentration, and also according to the coverage/vacancy pattern. 
It can also be noticed that in all cases, the individual titanium layers behave independently and the layer of carbon atoms forms a barrier between the individual layers. 
This makes it possible to work with only one side of the surface without disturbing the opposite side, allowing for better adjustment of the surface as required. 
We like to stress, that oxygen coverage between 30\% -- 100\% is also achieved experimentally.\cite{Persson2019} 
Therefore, despite DFT results are only the first step for the understanding of adsorption/desorption of terminal atoms (temperature effects, solvation, etc., should be investigated more deeply) and particular provided numbers (as coverage percentage for change of magnetism) should be considered with caution, DFT modeling can bring understanding on the atomic level and offer possibilities that can be targeted. 
\section{Conclusions}
In this work, we systematically investigated the structural and magnetic properties of \ce{Ti2C} MXene monolayers with various coverages and patterns of terminal oxygen and fluorine atoms, including partial, linear, and local defects. 
We have shown that terminal atom type affects both the pattern of the adsorption/vacancy defect and, subsequently, the magnetic properties of MXenes. 
In the case of adsorption by oxygen atoms, a linear zigzag pattern was found to be the most likely. 
That leads to a tuning of the magnetism from antiferromagnetic (AFM) and ferrimagnetic (FiM) spin arrangements to ferromagnetism (FM), depending on the number of adsorbed oxygen atoms. 
On the contrary, in the case of adsorption of fluorine atoms, a local pattern was found as preferred, with increasing magnetic moment. 
This results in a change of the FiM ground magnetic state to the FM state at a certain surface coverage (around 30\%). 
In the case of vacancy defect (or desorption) studies, we found the preferred pattern to be partial in the case of the O-vacancy and local in the case of the F-vacancy. 
The partial O-vacancy pattern has the consequence that the nonmagnetic character of \ce{Ti2CO2} remains up to ca. 20\% defective surface (around 80\% surface coverage). 
Subsequently, at lower surface coverages (\textit{c}~= 44\% –- 72\%), a change from NM to FM occurs. 
However, with the change of O-defect patterning to local or linear, spin polarization appears at the vacancy site, and a weak magnetic moment (2~$\mu_B$) is produced already at 8\% vacancy defect concentration. 
In the F-vacancy case, again, the local pattern with increasing magnetic moment is observed to be the most favorable with FM-arranged spin with a high total magnetic moment up to 32 $\mu_B$. 
Thus, the results show that the magnetic properties of \ce{Ti2CT2} MXenes are dependent on the electronegativity of the functional T group and can exhibit either FM, AFM, or FiM behavior. 
First, in the case of functional groups with a formal charge of -1$e$ (F), the FM ground state is predominant, whereas, in the case of -2$e$ (O), the ground state is significantly dependent on the adsorption/vacancy pattern. 
In all cases, it is also possible to observe that the individual titanium layers behave independently, and the carbon layer works as a barrier. 
That may represent the possibility of controlling only one layer of the material surface. 
This work provides encouraging results for the possibility of tunable magnetic behavior in \ce{Ti2CT2} MXenes using different patterns of adsorption/vacancy and terminal atoms. 
All presented results can provide promising guidance for modeling the functionalization of MXenes to obtain the desired properties of 2D materials. 


\section*{Conflicts of interest}
There are no conflicts to declare.

\section*{Acknowledgements}
This article has been produced with the financial support of 
the Czech Science Foundation (number 21-28709S)
and the European Union under the LERCO project (number CZ.10.03.01/00/22$\_$003/0000003) via the Operational Programme Just Transition. 
The computations were performed at IT4Innovations National Supercomputing Center through the e-INFRA CZ (ID:90140).



\balance


\bibliography{rsc} 
\bibliographystyle{rsc} 

\end{document}